\documentclass[aps,pra,10pt,notitlepage,nofootinbib,superscriptaddress,showkeys,showpacs]{revtex4-1}
\usepackage{amstext,amsmath,amssymb,amsfonts,bbm}
\usepackage[latin1]{inputenc}
\usepackage{graphicx}
\usepackage{hyperref}
\usepackage{epstopdf}
\usepackage{amsthm}
\usepackage{epsfig}
\usepackage{subfigure}
\usepackage{mathrsfs}

\newcommand\ph{\ensuremath{\varphi}}

\newcommand\ex[1]{e^{#1}}
\renewcommand\i{\ensuremath{\mathrm{i}}}

\newcommand\e[1]{_{\text{#1}}}

\renewcommand\lim[2]{\underset{ #1 \rightarrow #2 }{ \mathrm{lim} } \,}

\newcommand\pa[1]{\left( #1 \right)}
\newcommand\pac[1]{\left[ #1 \right]}

\newcommand\bra[1]{\left\langle \, \right|}

\newenvironment{systeme}
{ \left\{ \begin{aligned} }
{ \end{aligned} \right. }

\newcommand\define{\overset{\text{def.}}{=}}

\newcommand\wigner[1]{\left\{ \begin{matrix} #1 \end{matrix} \right\}}
\newcommand{\cc}{\text{c.c.}}
\newcommand{\vect}{\mathbf}
\newcommand{\tet}[1]{\ensuremath{\text{tet}_{#1}}}
\newcommand{\setEdm}{\ensuremath{\mathscr{E}}}
\newcommand{\setPR}{\ensuremath{\mathscr{P}}}
\newcommand{\intset}[2]{\ensuremath{\paac{#1,\ldots,#2}}}

\usepackage{color}

\def\beq{\begin{equation}}
\def\be{\begin{equation}}
\def\ee{\end{equation}}
\def\bes{\begin{eqnarray}}
\def\ees{\end{eqnarray}}

\DeclareMathOperator{\SU}{SU}
\DeclareMathOperator{\SO}{SO}

\DeclareMathOperator{\di}{d}

\def\bra{\langle}

\def\N{{\mathbbm N}}

\begin{document}

	
\title{\large \bf Asymptotics of Wigner 3nj-symbols with Small and Large Angular Momenta: an Elementary Method}
\author{{\bf Valentin Bonzom}}\email{vbonzom@perimeterinstitute.ca}
\author{{\bf Pierre Fleury}}\email{pierre.fleury@ens-lyon.fr}
\affiliation{Perimeter Institute for Theoretical Physics, 31 Caroline St. N, ON N2L 2Y5, Waterloo, Canada}
\date\today


\begin{abstract}
\noindent Yu and Littlejohn recently studied in arXiv:1104.1499 some asymptotics of Wigner symbols with some small and large angular momenta. They found that in this regime the essential information is captured by the geometry of a tetrahedron, and gave new formulae for 9j, 12j and 15j-symbols. We present here an alternative derivation which leads to a simpler formula, based on the use of the Ponzano-Regge formula for the relevant tetrahedron. The approach is generalized to Wigner 3nj-symbols with some large and small angular momenta, where more than one tetrahedron is needed, leading to new asymptotics for Wigner 3nj-symbols. As an illustration, we present 15j-symbols with one, two and four small angular momenta, and give an alternative formula to Yu's recent 15j-symbol with three small spins.
\end{abstract}

\maketitle

\section*{Introduction}

Wigner symbols are re-coupling coefficients of $\SU(2)$ representation theory. As such they naturally arise when dealing with sums of more than four spins and/or angular momenta in quantum mechanics, and are notoriously important in spectroscopy and atomic/molecular physics. Moreover, they also enter different parts of physics \cite{spinnets-marzuoli}, like quantum computing \cite{quantum-tet-marzuoli} (in the presence of topological order, and mainly with a quantum group deformation \cite{levin-wen-condensation}), and are expected to contain relevant aspects of quantum geometry (in loop quantum gravity and spin foam models \cite{PR1,bf-spinfoam-baez,noui-perez-ps3d,sf-action-fk}).

Although those objects are easily defined, using sums of Clebsch-Gordan coefficients, or inner products of wave-functions, it is quite hard to extract their semi-classical behavior, for large angular momenta (not even to mention getting a rigorous proof). Moreover, in the typical case of spin-orbit couplings, one may be interested into large angular momenta coupled to some intrinsic spins which cannot be scaled. Obviously, different behaviors are expected in those different regimes of a given symbol.

The most studied Wigner symbol is the 6j-symbol, whose asymptotics is known when all spins are large (the Ponzano-Regge formula, \cite{schulten-gordon2, roberts, louapre6j, razvan6j}), or with one being small (the Edmonds' formula \cite{edmonds, flude}). Over the years, the 6j-symbol has turned out to be an interesting topic in modern physics, see \cite{qm6j} which contains additional references. Very recent progress include sub-leading corrections to its Ponzano-Regge asymptotics \cite{6jnlo, 6jmaite, pushing6j}, a new recursion relation on the square of the 6j-symbol \cite{yetanother} and a derivation of its standard recursion relation as a Wheeler-DeWitt equation for three-dimensional Riemannian gravity \cite{3d-wdw}.

When all angular momenta are large, a nice feature of the Ponzano-Regge asymptotics is that all the information is captured into a tetrahedron whose edge lengths are basically the six quantum angular momenta. Such a simple geometric interpretation is usually not available for larger symbols, which makes their analysis more difficult. While a new approach was recently proposed in \cite{3nj-marzuoli} to classify the various asymptotic regimes of Wigner 3nj-symbols, it remains an open issue. In contrast with the generic case, it is interesting to note that the 15j-symbol admits a natural four-dimensional interpretation in the coherent state basis, in term of the geometry of a 4-simplex \cite{barrett-asym15j, barrett-asym-summary} (the same method has been applied to the three-dimensional Ponzano-Regge model, see \cite{dowdall-handlebodies}). Just like for the 6j-symbol, that geometric property can be understood via the recursions satisfied by the 15j-symbol which naturally arise as Hamiltonian equations for the quantum 4-simplex \cite{semiclass-paper}.

Littlejohn and Yu \cite{Yu_and_Littlejohn} have recently obtained a new formula for the asymptotics of the 9j-symbol with eight large angular momenta and one small spin (and also for larger symbols, but the method there focuses on this example, and we will do the same here). Their method is quite generic and powerful since it relies on previous works of the authors which extend the Born-Oppenheimer approximation in the case the fast degrees of freedom are coupled through a matrix of non-commuting operators. We refer the reader to \cite{Yu_and_Littlejohn} for further references to the method and its applications, as we will be more interested in the asymptotics of the 9j-symbol itself.

The asymptotic formula itself is indeed quite interesting. The information is encoded into the geometry of a tetrahedron, and can be re-formulated more conveniently with three tetrahedra. The formula also displays ingredients which are familiar to the asymptotics of the 6j-symbol, as noticed by the authors of \cite{Yu_and_Littlejohn}, more precisely the amplitude involving the inverse of the square-root of the volume of the tetrahedron, and oscillations with part of the frequency given by the Regge action of the same tetrahedron.

This suggests that the asymptotic formula for the 9j-symbol with a small spin may really be derived using the Ponzano-Regge asymptotic formula for some 6j-symbol associated to the relevant tetrahedron. This is exactly what we show in the present paper, ending up with a quite straightforward derivation. Remarkably, not only the derivation but also the final formula turns out to be simpler than that presented in \cite{Yu_and_Littlejohn}. The Section \ref{sec:9j} is devoted to showing this.

Our analysis further reveals the conditions so that an arbitrary Wigner symbol can be semi-classically described by a number of tetrahedra, when some spins remain small. This way we obtain generic asymptotic formulae for Wigner symbols, presented in the Section \ref{sec:3nj}.

As an illustration, we show the formulae for 15j-symbols in the Section \ref{sec:15j} since they may have some interesting applications in four-dimensional models for gravity. The cases with one, two and four small angular momenta are new. The case with three small angular momenta is an alternative to the recent result of \cite{yu15j} and our formula is a bit simpler since all quantities can be evaluated using only two tetrahedra. It should be noted that the method used in \cite{Yu_and_Littlejohn, yu12j, yu15j} is surely quite powerful since it has given access to some regimes which cannot be probed with our method. So at the end of the day, we think that both methods yield complementary results, with some overlap, and open an interesting window towards new asymptotics of re-coupling coefficients.

\section{Asymptotics of the 9j-symbol with one small angular momentum}
\label{sec:9j}

\subsection{Notations} \label{sec:conventions}

We have tried to use as often as possible the same notations as \cite{Yu_and_Littlejohn}. The large angular momenta, or spins in the sense of irreducible $\SU(2)$ representations, are denoted like $j\in\N/2$, and the small spins like $s\in\N/2$.

However, differences appear regarding angles. We systematically call $\ph_{a,b}$ the (internal) angle between two edges $a$, $b$ of a triangle. Its value is given in terms of the lengths of the triangle,
\be
\cos\ph_{a,b} = \frac{\ell_a^2 + \ell_b^2 - \ell_c^2}{2\ \ell_a\,\ell_b},
\ee
where the third length $\ell_c$ will be mentioned when necessary.

Lengths are simple functions of spins which will be attached to the corresponding edges,
\be
\ell_a \define j_a+\frac12 \define d_{j_a}/2.
\ee
This is the relation which is necessary for the Ponzano-Regge formula to work. But note that when using Edmonds' formula (given explicitly later), those dihedral angles are more naturally given in terms of different lengths $\sqrt{j_a(j_a+1)}$. The latter are asymptotically equivalent to $\ell_a$ and the difference in Edmonds' formula only appears at sub-leading orders. Note also that $d_j$ is the dimension of the representation of spin $j$.

We denote respectively $\Theta_e$ and $\theta_e$ the external and internal dihedral angles between two triangles adjacent to the edge $e$ in a tetrahedron, with $\Theta_e = \pi - \theta_e$. They are obviously determined by the lengths, and satisfy the following relation
\be \label{2d-3d_angles}
\cos\theta_a = \frac{\cos\ph_{b,c} \,-\,\cos\ph_{a,b}\,\cos\ph_{a,c} }{\sin\ph_{a,b}\,\sin\ph_{a,c}},
\ee
when the edges $a,b,c$ meet at a node in a tetrahedron.

\subsection{The formula}

The 9j-symbol comes as a recoupling coefficient, i.e. a change of basis, when describing in different ways the rotational invariant subspace of a tensor product of five spins\footnote{It is actually equivalent to look at the projection of $j_1\otimes j_2\otimes s\otimes j_4$ onto the spin $j_5$.}, say $j_1\otimes j_2\otimes s\otimes j_4\otimes j_5$. The invariant subspace is characterized by the fact that the sum of the angular momenta vanishes,
\be
\vect J_1 + \vect J_2 +\vect S + \vect J_4 + \vect J_5 = \vect{0} .
\ee
A first basis is obtained by choosing a spin $j_{13}$ in the tensor product $j_1\otimes s$, a spin $j_{24}$ in $j_2\otimes j_4$, and then consider the unique (normalized) vector satisfying $\vect J_{13} + \vect J_{24} + \vect J_5 = \vect 0$ in $j_{13}\otimes j_{24}\otimes j_5$. Denote this vector $\vert (j_1,s,j_{13}),j_5,(j_2,j_4,j_{24})\rangle$. An equivalent basis is obtained by tensoring first $j_1$ with $j_2$ and choosing a spin $j_{12}$ in the decomposition, and similarly choosing a spin $j_{34}$ in $s\otimes j_4$. A basis vector is then formed by the unique vector satisfying $\vect J_{12}+\vect J_{34}+\vect J_5 = \vect 0$, and is denoted $\vert (j_1,j_2,j_{12}),j_5,(s,j_4,j_{34})\rangle$. The 9j-symbol is just
\begin{equation}
\langle (j_1,j_2,j_{12}),j_5,(s,j_4,j_{34})\,\vert\,(j_1,s,j_{13}),j_5,(j_2,j_4,j_{24})\rangle
= \bigl[d_{j_{12}}\,d_{j_{34}}\,d_{j_{13}}\,d_{j_{24}}\bigr]^{\frac12} \begin{Bmatrix} j_1 &j_2 &j_{12}\\ s &j_4 &j_{34}\\ j_{13} &j_{24} &j_5\end{Bmatrix}
 ,
\end{equation}
where the normalization is a product of dimensions, $d_j \define 2j+1$.

The asymptotics when all spins are homogeneously scaled is given for instance in \cite{Varshalovich}.

\begin{figure}[!ht]
\centering
\includegraphics[height=3cm]{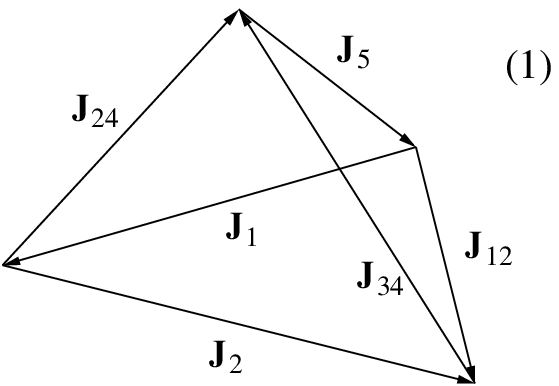} \hspace{2cm} \includegraphics[height=3cm]{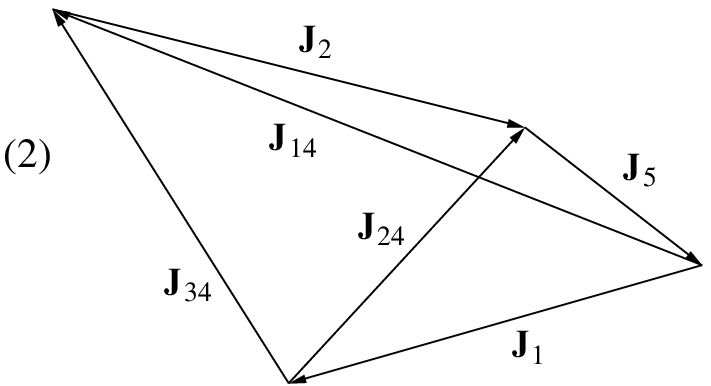}
\caption{Tetrahedron \tet{1} on the left, constructed with the six lengths $\{j_e + 1/2\}_{e=1,2,34,5,12,24}$, whose faces are the triads $(1,2,12)$, $(2,34,24)$, $(34,5,12)$ and $(1,5,24)$. We call $\vect{J}_e$ the vector associated with the edge $e$. Their orientations are such that $\vect{J}_{12} = \vect{J}_1 + \vect{J}_2$, $\vect{J}_{24} = \vect{J}_2 + \vect{J}_{34}$, and $\vect{J}_{34} + \vect{J}_5 + \vect{J}_{12} = \vect{J}_1 + \vect{J}_5 + \vect{J}_{24} = \vect{0}$. Tetrahedron \tet{2} on the right is constructed from the vectors $\vect{J}_1$, $\vect{J}_2$, $\vect{J}_{34}$, $\vect{J}_5$, $\vect{J}_{24}$. The last vector $\vect{J}_{14}$ is simply the sum $\vect{J}_1+\vect{J}_{34}$..}
\label{fig:tetrahedra_1_2}
\end{figure}

The result of \cite{Yu_and_Littlejohn} is an asymptotic formula for the 9j-symbol with one spin, $s$, being small (i.e. not scaled) while the other eight spins are homogeneously scaled by a large number. We will prove here an equivalent but different formula in that case. While the relevant quantities in the formula of \cite{Yu_and_Littlejohn} need three tetrahedra, all of them are here contained in only two tetrahedra. It is the following
\begin{equation}
\wigner{j_1 & j_2 & j_{12} \\
		s & j_4 & j_{34} \\
		j_{13} & j_{24} & j_5 \\}
\approx \frac{ (-1)^{j_{13} + j_2 + j_{34} + j_5 + s} }
			{ \sqrt{ d_{j_{1}}\, d_{j_{34}}\, (12 \pi V_1) } }
		\cos\pac{ \sum_{e \subset \text{tet}_1} \pa{j_e + \frac{1}{2}} \Theta_e^{(1)} + \frac{\pi}{4} - \mu \bigl(\pi- \theta_1^{(2)}\bigr) - \nu\, \theta_{34}^{(2)} }\
		\di_{\mu \nu}^{(s)} (\pi - \ph_{1,34}) .
\label{eq:main_result}
\end{equation}
In the above expression,
\begin{itemize}
\item the sum in the cosine runs over the six large angular momenta $\{j_e\}_e$ with $e=1, 2, 34, 5, 12, 24$ ;

\item $\mu \define j_{13} - j_1$ and $\nu \define j_{34} - j_4$, those two differences have the same order of magnitude as $s$, due to the triangle inequalities (or Clebsch-Gordan conditions)  between the spins $(j_1,s,j_{13})$ and  also between $(s,j_4,j_{34})$;

\item $V_1$ and $\{\Theta_e^{(1)}\}_{e=1,2,34,5,12,24}$ are geometric quantities associated to tetrahedron \tet{1} in Figure \ref{fig:tetrahedra_1_2}, $V_1$ is the volume of the tetrahedron and $\Theta_e^{(1)}$ is the angle between two external normals to the faces adjacent to the edge $e$ (external dihedral angle);

\item $\ph_{1,34}$ is associated with the tetrahedron \tet{2} given in the Figure \ref{fig:tetrahedra_1_2}, it is the angle between the edges $1$ and $34$; $\theta_1^{(2)}$ and $\theta_4^{(2)}$ are respectively the internal dihedral angles between the faces adjacent to the edges $1$ and $34$ in \tet{2}. 

\item $\di^{(s)}$ is the Wigner $d$-matrix with spin $s$, with the convention $\di^{(s)}_{\mu\nu}(\phi)= \langle s,
\mu\vert e^{-\frac{\i}2 \phi\sigma_y}\vert s,\nu\rangle$.
\end{itemize}

The formula holds in the classically allowed region away from the caustic, i.e. where the volume $V_1$ of the tetrahedron \tet{1} is not close to zero.

\begin{figure}[!ht]
\centering
\includegraphics[scale=0.8]{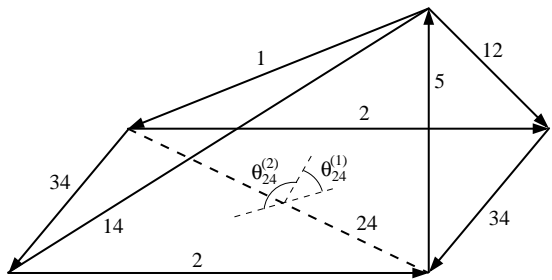}
\caption{That is a unified view of the two relevant tetrahedra \tet{1} and \tet{2}. They form a pyramid with a parallelogram basis spanned by $\vect J_2,\vect J_{34}$. Without using the vectors, one can define \tet{2} by setting the angle $\theta^{(2)}_{34} = \pi-\theta^{(1)}_{34}$, which is obvious in the picture, and that makes it possible to evaluate the additional length $j_{14}$.}
\label{fig:parall}
\end{figure}

The tetrahedron \tet{2} is built by gluing the triangles $(2,34,24)$ and $(1,5,24)$ along $24$, similarly to tetrahedron \tet{1}, but after flipping one of the triangles. This means that $2$ and $5$ meet at one end of $24$, and $1$ and $34$ meet at the other node. To completely determine \tet{2}, one has to set the dihedral angle between the two triangles. It is $\theta^{(2)}_{24} = \Theta^{(1)}_{24}$. This can be deduced, like in \cite{Yu_and_Littlejohn}, from the vectors representing the classical angular momenta, which are drawn on the Figure  \ref{fig:tetrahedra_1_2}. This is indeed equivalent to adding to \tet{1} the parallelogram spanned by $\vect J_2,\vect J_{34}$, and observe that \tet{2} is the tetrahedron between the parallelogram and \tet{1}. That observation is summarized in the Figure \ref{fig:parall} which offers a full view on the two relevant tetrahedra.

Our tetrahedra are closely related to those used in \cite{Yu_and_Littlejohn}. Their first tetrahedron is like ours but with $\vect{J}_{4}$ instead of $\vect{J}_{34}$ (notice indeed that they differ only by a vector of order $s$). Their second tetrahedron is built just like \tet{2} from \tet{1}. Because we are using a different first tetrahedron, the Regge action $\sum_{e \subset \text{tet}_1} \pa{j_e + \frac{1}{2}} \Theta_e^{(1)}$ in the cosine is not the same. The difference, which can be observed in \cite{Yu_and_Littlejohn}, is a term $\nu\Theta^{(1)}_{34}$. This adds to the explicit $\nu$ contribution in \eqref{eq:main_result}, which becomes $\nu(\Theta^{(1)}_{34}-\theta^{(2)}_{34})$. That new angle can be interpreted with the introduction of a third tetrahedron, which is the additional tetrahedron of \cite{Yu_and_Littlejohn}.

\subsection{Direct derivation of the formula}

Our derivation starts from a decomposition\footnote{There are several such decompositions. Those we are interested in are such that the summed variable, here $x$, has to satisfy Clebsch-Gordan conditions with the small spin $s$. There are four such possibilities. In our choice, $j_{24}$ will play a spcial role in the following. But we could have equally well chosen $j_2, j_{12}$ or $j_{34}$ instead.} of the 9j-symbol in terms of $6j$-symbols \cite{Varshalovich}
\begin{equation}
\wigner{j_1 & j_2 & j_{12} \\
	s & j_4 & j_{34} \\
	j_{13} & j_{24} & j_5}
=
\sum_x (-1)^{2x}\,d_x\,
		\wigner{j_1 & j_2 & j_{12} \\
			j_{34} & j_5 & x \\}
	\,\wigner{s & j_4 & j_{34} \\
			j_2 & x & j_{24} \\}
		\,\wigner{j_{13} & j_{24} & j_5 \\	
			x & j_1 & s \\}  .
\end{equation}
The range of summation of $x$ is finite due to triangle inequalities. $x$ is bounded from below by $\max(\vert j_{24}-s\vert, \vert j_1-j_5\vert, \vert j_2-j_{34}\vert)$, and from above by $\min(j_{24}+s, j_1+j_5, j_2+j_{34})$. Now the regime we are looking at, away from the caustic where the volume of \tet{1} becomes close to zero, ensures that neither $\max(\vert j_1-j_5\vert, \vert j_2-j_{34}\vert)$ nor $\min(j_1+j_5, j_2+j_{34})$ are close to $j_{24}$ up to terms of order $s$. Hence, from the non-degeneracy of \tet{1}, we known that $x$ runs from $j_{24}-s$ to $j_{24}+s$. Introduce $\xi \define x - j_{24}$ which lives in $\{-s,\dotsc,s\}$, and write $j_{13} = j_1+\mu, j_{34} = j_4+\nu$. Then,
\begin{equation}
\wigner{j_1 & j_2 & j_{12} \\
		s & j_4 & j_{34} \\
		j_{13} & j_{24} & j_5}
=
\sum_{\xi = -s}^{s} d_{j_{24}+\xi}\, (-1)^{2j_{24} + 2s}
		\wigner{s & j_{34}-\nu & j_{34} \\
			j_2 & j_{24} + \xi & j_{24} \\}
		\wigner{j_1 + \mu & j_{24} & j_5 \\
			j_{24} + \xi & j_1 & s \\}
	\wigner{j_1 & j_2 & j_{12} \\
			j_{34} & j_5 & j_{24} + \xi \\} .
\label{eq:decomposition}
\end{equation}

We observe that there are two different situations for the above 6j-symbols. The last one does not contain $s$, so that its large spin behavior is given by the Ponzano-Regge formula. The other two 6j-symbols only have five large spins, and their asymptotics are given by Edmonds' formula.

The presence of the symbol $\{\begin{smallmatrix} j_1 &j_2 &j_{12}\\ j_{34} &j_5 &j_{24}+\xi\end{smallmatrix}\}$ in \eqref{eq:decomposition} will make clear why the geometry of the tetrahedron \tet{1} is important. It is indeed well known that the asymptotics of that symbol is described using a tetrahedron geometry, with oscillations following the Regge action for that tetrahedron. Precisely,
\begin{equation}
\wigner{j_1 & j_2 & j_{12} \\
		j_{34} & j_5 & j_{24} + \xi \\}
\approx
\frac{\cos\pa{ S\e{R}(\xi) + \frac{\pi}{4} }}{\sqrt{12 \pi V(\xi)}} .
\end{equation}
$V(\xi)$ and $S\e{R}(\xi)$ are the volume and the Regge action of the tetrahedron constructed with the lengths $j_1 + 1/2$, $j_2 + 1/2$, $j_{12} + 1/2$, $j_{34} + 1/2$, $j_5 + 1/2$, $j_{24} + \xi + 1/2$, in such a way that the triangles are formed by the spins which are coupled in the $6j$-symbol (through Clebsch-Gordan couplings). Since that tetrahedron differs from the tetrahedron \tet{1} only due to a small $\xi$ of order $s$ along the vector $\vect J_{24}$, one can relate $S\e{R}(\xi)$ to the Regge action $S\e{R}^{(1)}$ of tetrahedron \tet{1} via
\begin{equation}
S\e{R}(\xi) \approx S\e{R}^{(1)} + \xi\, \Theta_{24}^{(1)} ,\quad \text{with}\qquad
S\e{R}^{(1)} \define \sum_{e \subset \text{tet}_1} \pa{j_e + \frac{1}{2}} \Theta_e^{(1)} .
\label{eq:relation_between_regge_actions}
\end{equation}
Here, the difference between the dihedral angles $\Theta_e(\xi)$ appearing in $S\e{R}(\xi)$ and the dihedral angles $\Theta^{(1)}_e$ of \tet{1} have been omitted thanks to the Schl\"{a}fli's identity (it makes sure that those sum to zero on the six edges of the tetrahedron). Besides, the variations of $V$ with $\xi$ are irrelevant at the leading order, so that $V(\xi) \approx V_1$.

The other two $6j$-symbols can be expressed using Edmonds' formula\footnote{It gives the asymptotic behavior of a $6j$-symbol when one spin is much smaller than the five others,
\begin{equation*}
\wigner{ a & b & c \\
		b+m & a+n & f }
\overset{a,b,c \gg m,n,f}{\approx}
\frac{ (-1)^{a+b+c+f+m} }
	{ \sqrt{(2a+1)(2b+1)} }
\di^{(f)}_{mn}(\ph_{a,b})  ,
\end{equation*}
where $\ph_{a,b}$ is the angle between the edges $a$ and $b$ in the triangle constructed with the three lengths $\ell_a$, $\ell_b$, $\ell_c$.
\label{fn:Edmonds}}
\cite{Varshalovich}:
\begin{gather}
\wigner{s & j_{34}-\nu & j_{34} \\
	j_2 & j_{24} + \xi & j_{24} \\}
\approx \frac{ (-1)^{j_2 + j_{24} + j_{34} + s} }
			{ \sqrt{d_{j_{34}}\,d_{j_{24}}} }
		\,\di_{-\xi \,-\nu}^{(s)}(\ph_{34,24})  ,\\[2mm]
\wigner{j_1 + \mu & j_{24} & j_5 \\
		j_{24} + \xi & j_1 & s \\}
\approx \frac{ (-1)^{j_{13} + j_{24} + j_5 } (-1)^{s+\xi} }
			{ \sqrt{d_{j_1}\,d_{j_{24}}} }
		\,\di_{\mu\,-\xi}^{(s)}(\ph_{1,24})  . 
\end{gather}
We use symmetries of the Wigner matrices to write $ \di_{-\xi \,-\nu}^{(s)}(\ph_{34,24}) = (-1)^{s -\xi} \di_{-\xi \nu}^{(s)}(\pi - \ph_{34,24}) $. Inserting those asymptotics in \eqref{eq:decomposition} and relabeling the sum by $\xi\mapsto -\xi$, we get
\begin{equation}
\{9j\}\approx
\frac{(-1)^{j_{13} + j_2 + j_{34} + j_5 + s}}
	{\sqrt{d_{j_1}\,d_{j_{34}}\,(12 \pi V_1)}}
\sum_{\xi = -s}^{s}
\cos\Bigl(S\e{R}^{(1)} + \frac{\pi}{4} - \xi \Theta_{24}^{(1)} \Bigr)
\ \di_{\mu\xi}^{(s)}(\ph_{1,24}) \,\di_{\xi\nu}^{(s)}(\pi-\ph_{34,24}) ,
\end{equation}

\begin{figure}[!ht]
\centering
\includegraphics[scale=0.8]{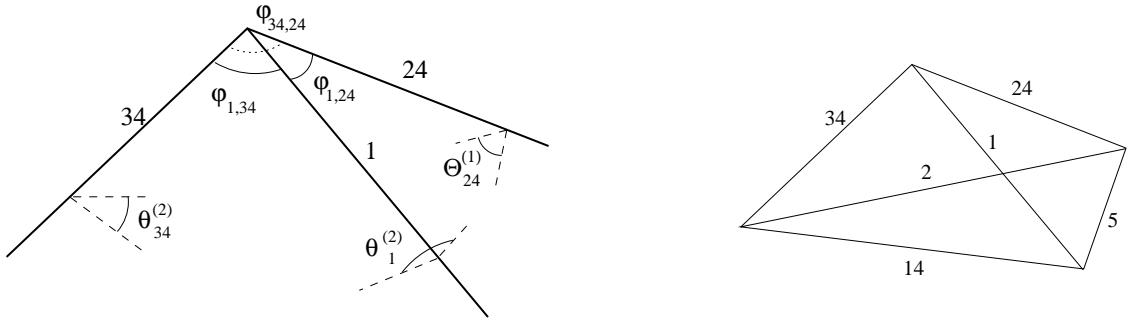}
\caption{Geometric interpretation of the Euler angles $\ph_{1,34}$, $\theta^{(2)}_1$ and $\theta^{(2)}_{34}$. They appear naturally as characteristics of tetrahedron \tet{2}, defined in Figure \ref{fig:tetrahedra_1_2}, and re-depicted here on the right.}
\label{fig:angles_one_node}
\end{figure}

The sum over $\xi$ is almost a matrix product between the two Wigner $d$-matrices, but not exactly because the argument of the cosine does depend on $\xi$. Nevertheless the sum over $\xi$ is a matrix product of Wigner $D$-matrices.
To see that, we simply write the cosine as a sum of exponentials. Let us call $A$ the prefactor of the sum. We get the following expression
\begin{equation}
\{9j\} \approx A\, \ex{\i (S_{\rm R}^{(1)}+\pi/4)} \sum_{\xi = -s}^{s}
\di_{\mu\xi}^{(s)}(\ph_{1,24}) \,\ex{ -\i\xi\Theta_{24}^{(1)} } \,\di_{\xi\nu}^{(s)}(\pi-\ph_{34,24})
+ \cc ,
\end{equation}
where `\cc' denotes the complex conjugate of the whole expression. By definition of the Wigner $D$-matrices, the sum can then be re-expressed as a matrix product,
\begin{equation}
\sum_{\xi} \di_{\mu\xi}^{(s)}(\ph_{1,24}) \,\ex{ -\i\xi\Theta_{24}^{(1)} } \,\di_{\xi\nu}^{(s)}(\pi-\ph_{34,24})
=  D_{\mu\nu}^{(s)} \pa{ \ex{ -\i \ph_{1,24} \frac{\sigma_y}{2} }
\,\ex{ -\i \Theta_{24}^{(1)} \frac{\sigma_z}{2} }
\,\ex{ - \i (\pi-\ph_{34,24}) \frac{\sigma_y}{2} } }.
\end{equation}
We have used above the notation in terms of $\SU(2)$ rotations, written with the Pauli matrices as generators. The last step of the calculation is to rewrite the relevant product of rotations as a single rotation parametrized by its Euler angles\footnote{Our convention for the Euler angles is the form $g=e^{-\frac{\i}2\alpha\sigma_z}e^{-\frac{\i}2\beta\sigma_y}e^{-\frac{\i}2\gamma\sigma_z}$ for $g\in\SU(2)$, together with $\sigma_z = \operatorname{diag}(1,-1), \sigma_y = \left(\begin{smallmatrix} 0&-\i\\ \i&0\end{smallmatrix}\right)$.}. That rewriting, from the angles $\ph_{1,24}, \Theta_{24}^{(1)}, \ph_{34,24}$ to Euler angles, will naturally be encoded into the geometry of the tetrahedron \tet{2}. Explicitly, we define three new angles $\theta^{(2)}_1, \ph_{1,34}, \theta^{(2)}_{34}$ by
\begin{equation}
\ex{ -\i \ph_{1,24} \frac{\sigma_y}{2} }
\,\ex{ -\i \Theta_{24}^{(1)} \frac{\sigma_z}{2} }
\,\ex{ - \i (\pi-\ph_{34,24}) \frac{\sigma_y}{2} }
=
\ex{-\i (\pi-\theta^{(2)}_1) \frac{\sigma_z}{2}}\,
\ex{-\i (\pi-\ph_{1,34}) \frac{\sigma_y}{2}}\,
\ex{-\i \theta^{(2)}_{34} \frac{\sigma_z}{2}}.
\end{equation}
The relation between both sets can be found in textbooks like \cite{Varshalovich}. Since all our angles lie in $[0,\pi]$, we can simply make use of the $\SO(3)$ relations,
\begin{gather}
\nonumber \cos\ph_{1,34} = \cos\ph_{1,24} \,\cos\ph_{34,24} + \sin\ph_{1,24} \,\sin\ph_{34,24} \,\cos\Theta_{24}^{(1)}  , \\[2mm]
\nonumber \cos\theta^{(2)}_{34} = \frac{\cos\ph_{1,24} - \cos\ph_{34,24}\,\cos\ph_{1,34}}{\sin\ph_{34,24}\,\sin\ph_{1,34}} ,\qquad  \cos\theta^{(2)}_{1} = \frac{\cos\ph_{34,24} - \cos\ph_{1,24}\,\cos\ph_{1,34}}{\sin\ph_{1,24}\,\sin\ph_{1,34}} ,
\\[2mm]
\frac{ \sin\theta^{(2)}_1 }
	{ \sin\ph_{34,24} }
=
\frac{ \sin\theta^{(2)}_{34} }
	{ \sin\ph_{1,24} }
=
\frac{ \sin\Theta_{24}^{(1)} }
	{ \sin\ph_{1,34} }  .
\label{eq:tet-angles}
\end{gather}
As our notation suggests, those new angles have a nice geometric interpretation with the help of the tetrahedron \tet{2}. The latter appears in the following way. Those equations characterize the relationship between 2d and 3d angles for three triangles forming the `top' of a tetrahedron, as displayed in Figure \ref{fig:angles_one_node}. The initial angles $\ph_{1,24},\Theta^{(1)}_{24},\ph_{34,24}$ enable to draw three links, $1,34,24$ meeting at a node, with two 2d angles being $\ph_{1,24}, \ph_{34,24}$ and the (internal) angle between the planes $(34,24)$ and $(1,24)$ being $\Theta^{(1)}_{24}$. Then the above formulae make it possible to evaluate the three remaining angles. The first equation states that $\ph_{1,34}$ is the angle between the edges $1$ and $34$. And $\theta^{(2)}_1, \theta^{(2)}_{34}$ given by the above formulae are the two other internal dihedral angles. Those geometric considerations are summarized in Figure \ref{fig:angles_one_node}. This is exactly the geometry of the `top' of \tet{2}, see Figure \ref{fig:tetrahedra_1_2}, where $1,34,24$ meet\footnote{Note that the edges $2$ and $5$ are somehow irrelevant here. This means that from the tetrahedron \tet{2} only the apex where $(1,34,24)$ meet is interesting, while the base triangle formed by $2,14,5$ does not provide interesting information.}.

The element $D_{\mu\nu}^{(s)}$ of the Wigner D-matrix of the relevant $\SU(2)$ rotation can then be written in term of the element $\di^{(s)}_{\mu\nu}$ of a $d$-matrix, so that the 9j-symbol takes the following form
\begin{equation}
\{9j\} \approx
A \exp\pac{ \i\pa{ S_{\rm R}^{(1)}+\frac\pi 4 - \mu\bigl(\pi-\theta_1^{(2)}\bigr) - \nu\theta_{34}^{(2)} } } \,\di^{(s)}_{\mu\nu}( \pi - \ph_{1,34} ) + \cc ,
\end{equation}
that is to say
\begin{equation}
\wigner{j_1 & j_2 & j_{12} \\
		s & j_4 & j_{34} \\
		j_{13} & j_{24} & j_5}
\approx
\frac{(-1)^{j_{13} + j_2 + j_{34} + j_5 + s}}
	{\sqrt{d_{j_1}\,d_{j_{34}}\,(12 \pi V_1)}}
\cos\pac{ S\e{R}^{(1)} + \frac{\pi}{4} - \mu\bigl(\pi-\theta_1^{(2)}\bigr) - \nu\theta_{34}^{(2)} }
\,\di^{(s)}_{\mu\nu}( \pi - \ph_{1,34} )  .
\end{equation}
This is exactly the formula \eqref{eq:main_result}.

\begin{figure}[!ht]
\centering
\subfigure[\label{fig:j24}]{\includegraphics[scale=0.75]{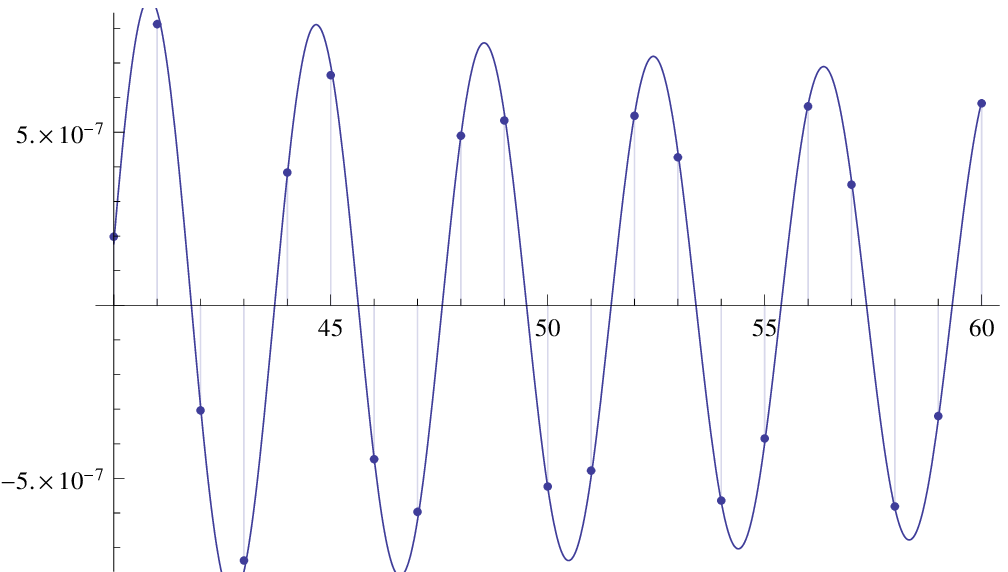}}
\hspace{1.5cm}
\subfigure[\label{fig:error}]
{  \includegraphics[scale=0.75]{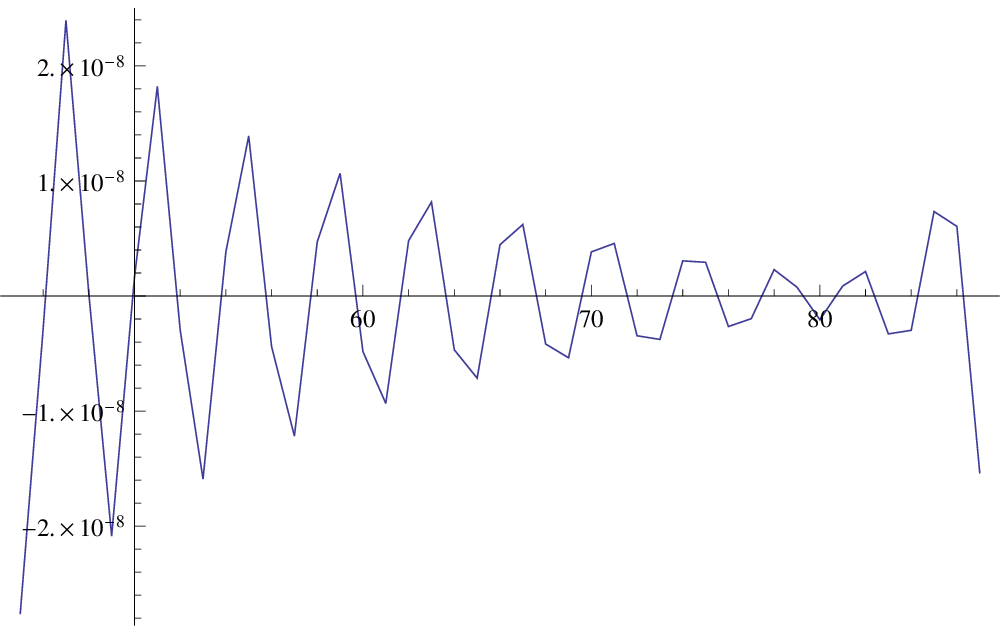}}
\subfigure[\label{fig:j1}]{
\includegraphics[scale=0.75]{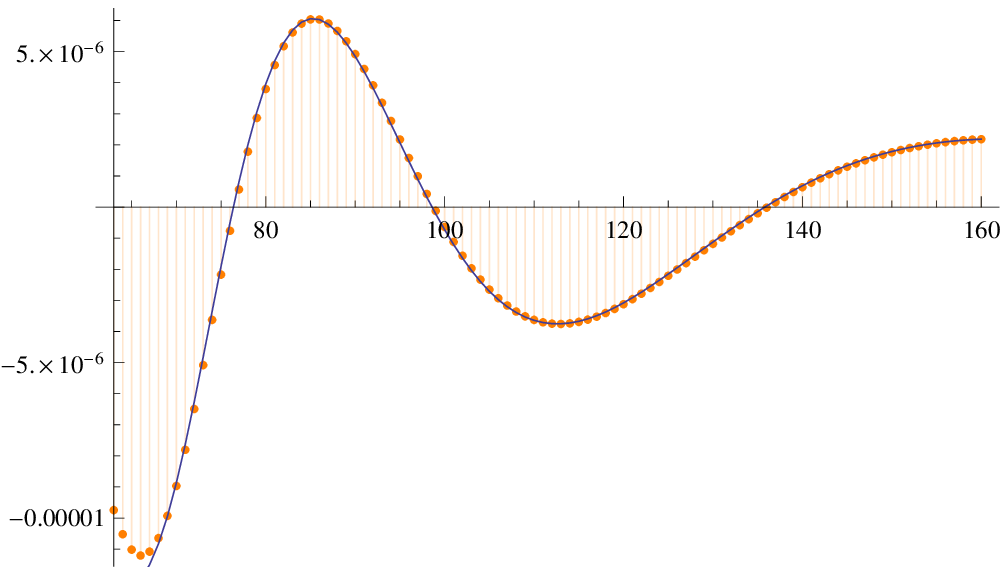}}
\hspace{1.5cm}
\subfigure[\label{fig:j5}]{
\includegraphics[scale=0.75]{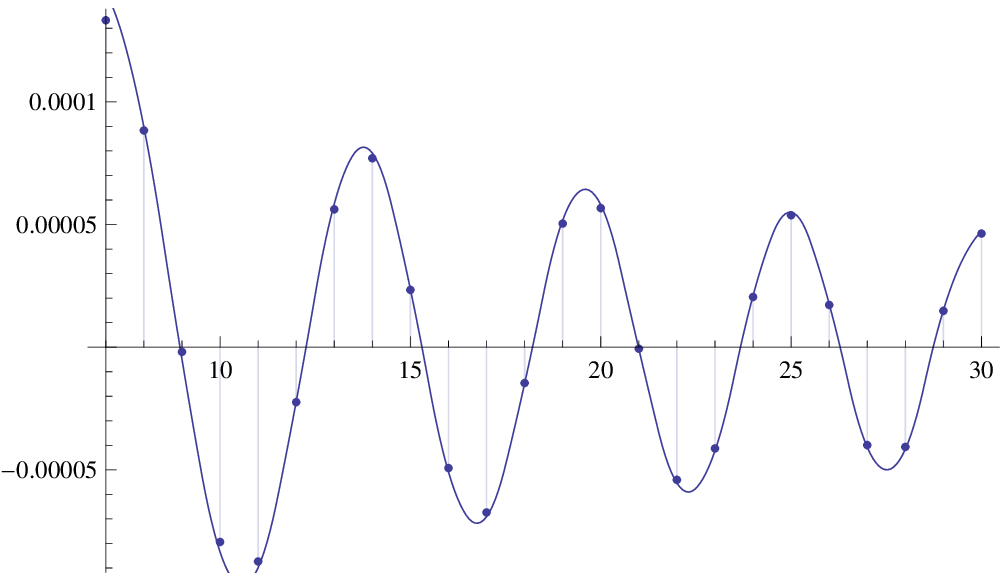}}

\caption{On \ref{fig:j24}, \ref{fig:j1}, \ref{fig:j5}, the points correspond to the exact 9j-symbols and the curve to the asymptotic formulae. Figure \ref{fig:j24} is the symbol $\left\{\begin{smallmatrix}430 &30 &430\\ 1 & 60 &61\\ 431 &j_{24}&430\end{smallmatrix}\right\}$ and \ref{fig:error} plots the absolute difference between the exact values and the asymptotics. Figure \ref{fig:j1} plots the symbol $\left\{\begin{smallmatrix}j_1+\frac{1}2 &\frac{201}2 &j_1+3\\ 1 & 60 &61\\ j_1+\frac32 &j_{24}&\frac{99}2\end{smallmatrix}\right\}$, and Figure \ref{fig:j5} the symbol $\left\{\begin{smallmatrix}\frac{51}2 &\frac{53}2 &28\\ \frac12 & \frac{47}2 &24\\ 25 &27&j_5\end{smallmatrix}\right\}$.}
\label{fig:numerics}
\end{figure}

Since our formula is different from that of Yu and Littlejohn, it is worth comparing it directly with numerics. From the derivation, it is clear that our formula holds as long as the Edmonds' and Ponzano-Regge formulae do. The comparison with the numerically computed 9j-symbol is indeed very good, as it can be seen on the plots of Figure \ref{fig:numerics}. To show that the agreement becomes better at large spins, we used the symbol $\left\{\begin{smallmatrix}j_1+\frac{1}2 &\frac{201}2 &j_1+3\\ 1 & 60 &61\\ j_1+\frac32 &j_{24}&\frac{99}2\end{smallmatrix}\right\}$, Figure \ref{fig:j1}, for values of $j_1$ from $63$ to $160$. For the smallest values of $j_1$, the approximation breaks down because the volume of \tet{1} becomes negative. The same phenomenon is observed on Figure \ref{fig:j24} and the error plot \ref{fig:error}, where the error increases when we reach the low and high values of $j_{24}$.

Notice that the corresponding symbol is $\left\{\begin{smallmatrix}430 &30 &430\\ 1 & 60 &61\\ 431 &j_{24}&430\end{smallmatrix}\right\}$, whose large spins differ by a factor $10$. It means that the approximation is still good when \tet{1} is distorted with some large angular momenta larger than other.

\section{Asymptotics of 3nj-symbols with small and large angular momenta} \label{sec:3nj}

\subsection{Decomposition of a 3nj-symbol}

For $n\geq4$, there exist several ways to define a 3nj-symbol. For instance, in the case $n=4$, one can define two different kinds of (irreducible) 12j-symbols. We restrict the discussion to the first kind, in the terminology of \cite{jucys}, whose decomposition in terms of 6j-symbols is
\begin{equation}
\wigner{j_1 & & j_2 & \ldots & j_n & \\
	& l_1 & & l_2 & \ldots & l_n \\
	k_1 & & k_2 & \ldots & k_n & }
=
\sum_x d_x(-1)^{R_n+(n-1)x}
\wigner{j_1 & k_1 & x \\
		k_2 & j_2 & l_1}
\wigner{j_2 & k_2 & x \\
		k_3 & j_3 & l_2}
 \dotsm\
\wigner{j_{n-1} & k_{n-1} & x \\
		k_n & j_n & l_{n-1}}
\wigner{j_n & k_n & x \\
		j_1 & k_1 & l_n} ,
\label{eq:decomposition_3nj}
\end{equation}
where $R_n \define \sum_{i=1}^n j_i + k_i + l_i$.
In Equation \eqref{eq:decomposition_3nj}, we wrote the 3nj-symbol so that the coupled spin triads appear easily, namely: $(j_1,l_1,j_2)$, $(j_2,l_2,j_3)$, \ldots, $(j_{n-1},l_{n-1},j_n)$, $(j_n,l_n,k_1)$, $(k_1,l_1,k_2)$, \ldots, $(k_{n-1},l_{n-1},k_n)$, $(k_n,l_n,j_1)$. Those couplings are also fully represented in the involved $6j$-symbols.

Symmetries of the 3nj-symbols can be deduced from the symmetries of the 6j-symbols in the above decomposition \cite{Varshalovich}. It is convenient to distinguish two groups of spins which have different roles, $(j_1,\dotsc,j_n,k_1,\dotsc,k_n)$ and $(l_1,\dotsc,l_n)$. Then the 3nj-symbol is invariant under simultaneous circular permutations within those groups. It is also invariant under the exchange of the $j$-row with the $k$-row.

\subsection{Hypotheses - applicability of the method}
\label{sbsec:hypotheses}

Almost all the angular momenta are large, and a few of them can be chosen not to scale. Our method applies when the following conditions are satisfied:
\begin{enumerate}
\item \emph{one and only one} spin among $j_1,\ldots,j_n,k_1,\ldots,k_n$ is small, every other small spin must be an $l_i$;
\item we are \emph{away from the caustic}, i.e. the volumes of the tetrahedra associated with the 6j-symbols in \eqref{eq:decomposition_3nj} which do only have large spins are far from zero.
\item the small $l_i$ must be chosen so that in each $6j$-symbol of the decomposition \eqref{eq:decomposition_3nj}, there is \emph{at most one} small spin;
\end{enumerate}
The two first conditions ensure that the summed variable $x$ in \eqref{eq:decomposition_3nj} is of the order of the large spins, while its range is controlled by the small spin chosen among $(j_1,\dotsc,j_n,k_1,\dotsc,k_n)$ \footnote{In fact, that explains only why we need \emph{at least} one small spin among $(j_1,\dotsc,j_n,k_1,\dotsc,k_n)$. The reason why we must have \emph{at most} one appears later. The main idea is to ensure that the sum over $x$ is a product of two Wigner matrices.}. Precisely, if $j_1$ is small, then $k_1-j_1 \leq x\leq k_1+j_1$. The last restriction enables us to use standard asymptotic expressions of the $6j$-symbol, the Ponzano-Regge and Edmonds' formulae.

\subsection{Asymptotics of the summand}

Without loss of generality (thanks to the symmetries of the symbol) we choose $j_1$ to be small. Let also $\{l_m\}_{m \in \setEdm}$ be small, for a set of integers $\setEdm \subset \intset{2}{n-1}$. We introduce the following half-integers which are small due to the triangular inequalities,
\begin{equation}
\begin{systeme} \xi &\define x - k_1 \\
				\mu &\define j_2 - l_1 \\
				\nu &\define k_n - l_n
\end{systeme}
\quad \text{and} \qquad
\forall m \in \setEdm \
\begin{systeme} \eta_m &\define j_{m+1} - j_m \\
				\kappa_m &\define k_{m+1} - k_m,
\end{systeme}
\end{equation}
where $\xi,\mu,\nu \in \intset{-j_1}{j_1}$ and $\forall m \in \setEdm \quad \eta_m,\kappa_m \in \intset{-l_m}{l_m}$. We are now ready to use asymptotic formulae for each $6j$-symbol involved in \eqref{eq:decomposition_3nj}.

\subsubsection{6j-symbols with one small spin}

\begin{figure}[!ht]
\centering
\includegraphics[scale=0.8]{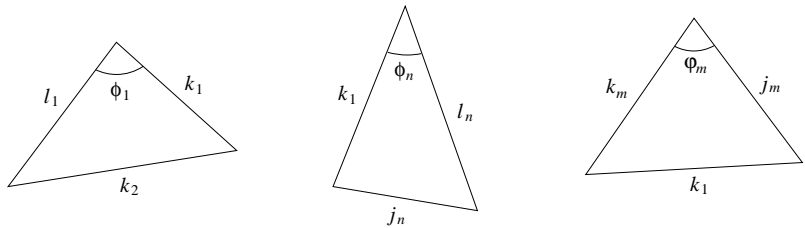}
\caption{Definitions of $\phi_{1}$, $\phi_n$ and $(\ph_m)_{m \in \setEdm}$ as angles in triangles. Each edge $e$ is labelled by a spin $j_e$ which indicates its length $\ell_e = \sqrt{j_e(j_e+1)} \approx j_e+1/2$.}
\label{fig:angles_3_triangles}
\end{figure}

Edmonds' formula applies to the 6j-symbols with one small spin,
\begin{align}
\wigner{j_1 & k_1 & x \\
		k_2 & j_2 & l_1}
&\approx
\frac{(-1)^{ (j_1+\mu) + (k_1 + k_2 + l_1) } }{\sqrt{d_{k_1}\,d_{l_1}}}\, \di^{(j_1)}_{\mu \xi}(\phi_1) , \\[2mm]
\wigner{j_n & k_n & x \\
		j_1 & k_1 & l_n}
&\approx
\frac{(-1)^{ (j_1+\xi) + (j_n + l_n + k_1) } }{\sqrt{d_{l_n}\,d_{k_1}}} \,\di^{(j_1)}_{\xi \nu}(\phi_n) ,
\end{align}
and $\forall m \in \setEdm$
\begin{equation}
\wigner{j_m & k_m & x \\
		k_{m+1} & j_{m+1} & l_m}
\approx
\frac{ (-1)^{(\kappa_m+l_m) + (j_m + k_m + k_1 + \xi) } }{ \sqrt{ d_{j_m}\,d_{k_m} } } \,\di^{(l_m)}_{\kappa_m \eta_m} (\ph_m).
\end{equation}
The angles $\phi_1$, $\phi_n$ and $(\ph_m)_{m \in \setEdm}$ are defined geometrically in Figure  \ref{fig:angles_3_triangles}. As we perform all the calculations to the leading order, we can neglect the variations of $\ph_m$ with $\xi$, namely we take $k_1$ instead of $k_1+\xi$ for the construction of the triangle $(j_m,k_m,k_1+\xi)$ in the Figure \ref{fig:angles_3_triangles}. Also note that $l_1 \approx j_2$ and $l_n \approx k_n$, because $j_1$ is small.

\subsubsection{6j-symbols with six large spins}

\begin{figure}[!ht]
\centering
\includegraphics[scale=0.8]{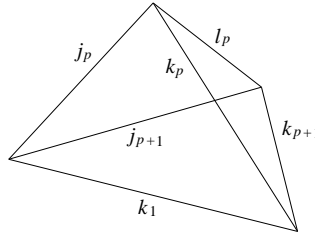}
\caption{Tetrahedron \tet{p} canonically associated with the $6j$-symbol $\left\{\begin{smallmatrix}j_p & k_p & k_1 \\ k_{p+1} & j_{p+1} & l_p\end{smallmatrix}\right\}$. Each edge $e$ is labeled by a spin $j_e$ which indicates its length $\ell_e = j_e+1/2$.}
\label{fig:tet_regge}
\end{figure}

Let $\setPR$ be the set of labels $p$ corresponding to $6j$-symbols with six large spins, namely $\setPR \define \intset{2}{n-1} \smallsetminus \setEdm$. We apply the Ponzano-Regge formula to those symbols, and use the same development of the Regge action as in Equation \eqref{eq:relation_between_regge_actions} to get for all $p \in \setPR$
\begin{equation}
\wigner{j_p & k_p & x \\
		k_{p+1} & j_{p+1} & l_p}
\approx
\frac{1}{\sqrt{12\pi V_p}} \cos\pa{ S\e{R}^{(p)} + \xi \Theta_{k_1}^{(p)} + \frac{\pi}{4} },
\label{eq:ponzano-regge}
\end{equation}
where $V_p$ and $S\e{R}^{(p)}$ are respectively the volume and the Regge action of the tetrahedron \tet{p} depicted in Figure \ref{fig:tet_regge}. It is constructed in the usual way from the $6j$-symbol in the left-hand side \eqref{eq:ponzano-regge}, for $\xi=0$.

The Regge action $S\e{R}^{(p)}$ of that tetrahedron is explicitly given by $S\e{R}^{(p)} \define \sum_{e \subset \text{tet}_p} (j_e + 1/2) \Theta_e^{(p)}$, where $\Theta_{e}^{(p)}$ is the external dihedral angle at the edge $e$ in \tet{p}.

\subsection{Evaluation of the sum}
\label{sbsec:reduction_sum}

Gathering the above pieces into the decomposition \eqref{eq:decomposition_3nj} of the 3nj-symbol, we obtain
\begin{multline}
\{3nj\}
\approx
\frac{(-1)^{r_n(\setEdm)} }{ \sqrt{d_{l_1}\,d_{l_n}} }
\pac{
\prod_{m\in\setEdm} \frac{ \di_{\kappa_m \eta_m}^{(l_m)}(\ph_m) }{ \sqrt{d_{j_m}d_{k_m}} }
}\\
\times \sum_{\xi=-j_1}^{j_1}
(-1)^{(n+M)(j_1-\xi)} \,\di^{(j_1)}_{\mu \xi}(\phi_1) \,\di^{(j_1)}_{\xi \nu}(\phi_n)
\prod_{p\in\setPR} \frac{1}{\sqrt{12\pi V_p}} \,\cos\pa{ S\e{R}^{(p)} + \xi \Theta_{k_1}^{(p)} + \frac{\pi}{4} } ,
\end{multline}
with $M \define \lvert\setEdm\rvert$ and $r_n(\setEdm) \define R_n + (n+M-1)(k_1+j_1) + (\mu-j_1) + (k_1+k_2+l_1) + (k_1+j_n+l_n) + \sum_{m\in\setEdm} j_m+l_m+k_{m+1}$. In practice, we expect that this asymptotic formula can be used numerically. Every angle which is involved has a precise definition in terms of the angular momenta $\{j_i,k_i,l_i\}_i$, and sums and products are finite.

But we can go one step further and recast the sum over $\xi$ as a $D$-matrix product. That leads to a general asymptotic formula for 3nj-symbols, which we do not think is more powerful in terms of numerical computations, but which has a clearer geometric meaning.

As in the case of the 9j-symbol, we extract $\xi$ from the argument of the cosines by writing each cosine as a sum of complex exponentials. Expanding the product of cosines and re-organizing its terms, we obtain the following combinatorial expression
\begin{equation}
\prod_{p\in\setPR} \cos\pa{ S\e{R}^{(p)} + \xi \Theta_{k_1}^{(p)} + \frac{\pi}{4} }
=
\frac{1}{2^{P+1}} \sum_{\substack{ \{\sigma_p = \pm 1,\\ p\in\setPR\} }}
\exp \ \i\,\biggl[\ \sum_{p\in\setPR} \sigma_p \pa{ S\e{R}^{(p)} + \frac{\pi}{4} } + \xi \sum_{p\in\setPR} \sigma_p\,\Theta_{k_1}^{ (p) } \biggr] + \cc ,
\label{eq:product_cos_symmetry}
\end{equation}
where $P \define \lvert \setPR\rvert $ is the number of Ponzano-Regge formulae which we have used. Note that in \eqref{eq:product_cos_symmetry} we have used the symmetry $(\sigma_p \rightarrow -\sigma_p)$ of the sum to make the complex conjugation explicit.

We perform the sum over $\xi$ for each configuration $\{\sigma\}$ independently. Since $j_1$ and $\xi$ may not be integers, we write $(-1)^{(n+M)(j_1-\xi)} = \exp \,\i\pi (n+M)(j_1-\xi)$ and define the angle
\begin{equation} \label{eq:defomega}
\omega^{\{\sigma\}}_{k_1} \define (n+M)\pi - \sum_{p\in\setPR} \sigma_p\Theta^{(p)}_{k_1} \pmod{4 \pi}.
\end{equation}
The reason why it is defined only modulo $4\pi$ is that generically $j_1\in\N/2$. The sum over $\xi$ reads
\begin{equation}
\sum_{\xi=-j_1}^{j_1}  \di^{(j_1)}_{\mu \xi}(\phi_1) \,\ex{ -\i \xi \omega_{k_1}^{ \{\sigma\} } } \,\di^{(j_1)}_{\xi \nu}(\phi_n)
= D^{(j_1)}_{\mu\nu}\bigl( e^{-\frac{\i}2 \phi_1\sigma_y}\ e^{-\frac{\i}2 \omega^{\{\sigma\}}_{k_1} \sigma_z}\ e^{-\frac{\i}2 \phi_n \sigma_y}\bigr).
\label{eq:matrix_product}
\end{equation}
Like in the case of the 9j-symbol, the final step is to find the Euler angles of the $\SU(2)$ rotation on the right hand side. They have a nice geometric picture as angles of a tetrahedron, but provided $\omega$ ranges in an interval of size $\pi$ (so that it can be a dihedral angle). Since $\omega^{\{\sigma\}}_{k_1}$ is generically in $[-2\pi,2\pi]$, we distinguish four cases.

\begin{figure}[!ht]
\centering
\includegraphics[scale=0.8]{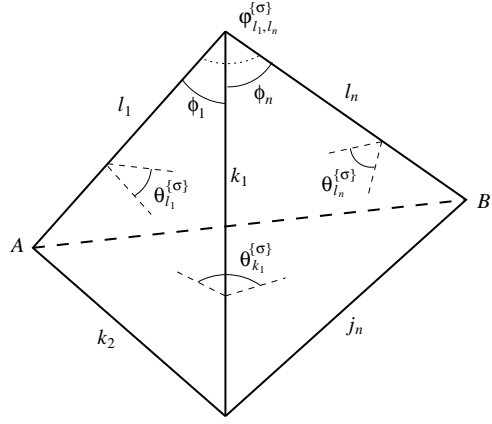}
\caption{Tetrahedron \tet{\{\sigma\}} associated with a sign configuration $\{\sigma\}$. It is determined by the triangles $(k_1, k_2, l_1)$ and $(j_n, l_n, k_1)$, and by the angle $\theta_{k_1}^{\{\sigma\}}$ between them, which in turn determine the length $AB$.}
\label{fig:tetrahedron_configuration_sigma}
\end{figure}

\begin{enumerate}
 \item \label{case1}{\bf Case $\omega^{\{\sigma\}}_{k_1} \in [0,\pi]$}. The formula for the Euler angles show that $\omega^{\{\sigma\}}_{k_1}$ has a natural interpretation as an {\em external} dihedral angle. So we consider the tetrahedron \tet{\{\sigma\}} depicted in Figure \ref{fig:tetrahedron_configuration_sigma}, defined by the gluing of the triangles $(k_1,k_n,j_n)$ and $(k_1,l_1,k_2)$ (which carry the angles $\phi_1$ and $\phi_n$, see Figure \ref{fig:angles_3_triangles}) with dihedral angle
\begin{equation}
\theta^{\{\sigma\}}_{k_1} \define \pi - \omega^{\{\sigma\}}_{k_1}.
\end{equation}
It is such that
\begin{equation}
D^{(j_1)}_{\mu\nu}\bigl( e^{-\frac{\i}2 \phi_1\sigma_y}\ e^{-\frac{\i}2 (\pi - \theta^{\{\sigma\}}_{k_1}) \sigma_z}\ e^{-\frac{\i}2 \phi_n \sigma_y}\bigr) = e^{-\i\mu \theta^{\{\sigma\}}_{l_1}}\ \di^{(j_1)}_{\mu\nu}(\ph^{\{\sigma\}}_{l_1,l_n})\ e^{-\i\nu \theta^{\{\sigma\}}_{l_n}},
\end{equation}
where the angles are shown on Figure \ref{fig:tetrahedron_configuration_sigma} and satisfy the relations \eqref{2d-3d_angles}, \eqref{eq:tet-angles}.

\item {\bf Case $\omega^{\{\sigma\}}_{k_1} \in [-2\pi,-\pi]$}. Then we write
\be
\theta^{\{\sigma\}}_{k_1} \define -\pi-\omega^{\{\sigma\}}_{k_1},
\ee
which is in $[0,\pi]$. Notice that
\be \label{eq:omega0pi}
e^{-\i\xi \omega^{\{\sigma\}}_{k_1}} = e^{2\i\pi\xi}\ e^{-\i(\pi-\theta^{\{\sigma\}}_{k_1})\xi} = e^{2\i\pi j_1}\ e^{-\i(\pi-\theta^{\{\sigma\}}_{k_1})\xi}.
\ee
Hence, we are back to the case \ref{case1}, and the final formula only differs by a phase $(-1)^{2j_1}$.

\item \label{case3}{\bf Case $\omega^{\{\sigma\}}_{k_1} \in [-\pi,0]$}. We will have again the tetrahedron \tet{\{\sigma\}} where the internal angle at $k_1$ is
\be
\theta^{\{\sigma\}}_{k_1} \define \pi+\omega^{\{\sigma\}}_{k_1},
\ee
Using the symmetries of Wigner matrices, we find that the relevant matrix element of the $\SU(2)$ rotation is
\be
D^{(j_1)}_{\mu\nu}\bigl( e^{-\frac{\i}2 \phi_1\sigma_y}\ e^{-\frac{\i}2 \omega^{\{\sigma\}}_{k_1} \sigma_z}\ e^{-\frac{\i}2 \phi_n \sigma_y}\bigr)
= e^{\i\pi j_1}\ D^{(j_1)}_{-\mu\nu}\Bigl( e^{-\frac{\i}2 \theta^{\{\sigma\}}_{l_1}\sigma_z}\, e^{-\frac{\i}2 (\pi-\ph^{\{\sigma\}}_{l_1,l_n})\sigma_y}\,e^{-\frac{\i}2 (\pi-\theta_{l_n}^{\{\sigma\}})\sigma_z}\Bigr).
\ee

\item {\bf Case $\omega^{\{\sigma\}}_{k_1} \in [\pi,2\pi]$}. The internal angle of \tet{\{\sigma\}} is
\be
\theta^{\{\sigma\}}_{k_1} \define -\pi+\omega^{\{\sigma\}}_{k_1}.
\ee
Following the reasoning of \eqref{eq:omega0pi}, we conclude that it differs from the case \ref{case3} only by a factor $(-1)^{2j_1}$.

\end{enumerate}

\subsection{Final asymptotics formula}
\label{sbsec:final}

The above cases fit into a not-so-complicated formula.
\begin{multline}
\label{eq:general_formula}
\{3nj\}
\approx\,
\frac{ (-1)^{r_n(\setEdm)} }{2^{{P}}\sqrt{d_{l_1}\,d_{l_n}}}
\, \pac{
\prod_{p\in\setPR} \frac{1}{\sqrt{12\pi V_p}}
\ \prod_{m\in\setEdm} \frac{ \di_{\kappa_m \eta_m}^{(l_m)}(\ph_m) }{ \sqrt{d_{j_m}\,d_{k_m}} }
}\\
\times \sum_{ \{\sigma_p = \pm 1\}_{p\in\setPR} }
\cos\biggl[\ \sum_{p\in\setPR}\sigma_p\Bigl(S_{\rm R}^{(p)}+\frac{\pi}4\Bigr) + \pi (n+M)j_1 +f^{\{\sigma\}}_{\mu\nu}\biggr]
\ \di_{\mu\nu}^{(j_1)} \pa{ \phi_{l_1,l_n}^{\{\sigma\}} } .
\end{multline}

\begin{itemize}
\item $\setPR$ is the set of tetrahedra on which the Ponzano-Regge formula has been applied, depicted in Figure \ref{fig:tet_regge}, and $P=\lvert \setPR\rvert$. $\setEdm$ is the subset of $\{2,\dotsc,n-1\}$ corresponding to the small spins $l_m$, with $M=\lvert\setEdm\rvert$. The global sign is given by $r_n(\setEdm) = R_n + (n+M-1)(k_1+j_1) + (\mu-j_1) + (k_1+k_2+l_1) + (k_1+j_n+l_n) + \sum_{m\in\setEdm} j_m+l_m+k_{m+1}$.

\item $\mu=j_2-l_1$ and $\nu=k_n-l_n$ are small, of the order of magnitude of $j_1$. Also $\kappa_m = k_{m+1}-k_m, \eta_m=j_{m+1}-j_m$ are of order $l_m$ for $m\in\setEdm$.

\item $V_p$ and $S_{\rm R}^{(p)}$ are the volumes and Regge actions of the tetrahedra \tet{p}, depicted in the Figure \ref{fig:tet_regge} and which have only large spins.

\item To each sign configuration $\{\sigma\}$, we have assigned a tetrahedron \tet{\{\sigma\}}, Figure \ref{fig:tetrahedron_configuration_sigma}, defined by the gluing of the two triangles along $k_1$ with a dihedral angle $\theta^{\{\sigma\}}_{k_1}$ given in term of $\omega^{\{\sigma\}}_{k_1} = (n+M)\pi-\sum_{p\in\setPR} \sigma_p\Theta^{(p)}_{k_1}$ in the above subsection. The function $f_{\mu\nu}^{\{\sigma\}}$ depends on the value of $\omega^{\{\sigma\}}_{k_1}$ as follows,
\be
f_{\mu\nu}^{\{\sigma\}} = \left\{
\begin{aligned}
 &-& &(\mu\theta^{\{\sigma\}}_{l_1} + \nu\theta^{\{\sigma\}}_{l_n})& +\,2\pi j_1\quad & \text{if}\ \omega^{\{\sigma\}}_{k_1}\in[-2\pi,-\pi],\\
 & & &(\mu\theta^{\{\sigma\}}_{l_1} + \nu\theta^{\{\sigma\}}_{l_n})& \qquad & \text{if}\ \omega^{\{\sigma\}}_{k_1}\in[-\pi,0],\\
 &-& &(\mu\theta^{\{\sigma\}}_{l_1} + \nu\theta^{\{\sigma\}}_{l_n})& \qquad & \text{if}\ \omega^{\{\sigma\}}_{k_1}\in[0,\pi],\\
 & & &(\mu\theta^{\{\sigma\}}_{l_1} + \nu\theta^{\{\sigma\}}_{l_n})& +\,2\pi j_1\quad & \text{if}\ \omega^{\{\sigma\}}_{k_1}\in[\pi,2\pi].\\
\end{aligned}\right.
\ee
The angles $\ph^{\{\sigma\}}_{l_1,l_n}, \theta^{\{\sigma\}}_{l_1}, \theta^{\{\sigma\}}_{l_n}$ are evaluated from the angles $\phi_1,\phi_n, \theta^{\{\sigma\}}_{k_1}$ following the Figure \ref{fig:tetrahedron_configuration_sigma}.
\end{itemize}

In our final result, there is a remaining sum over sign assignments $\{\sigma\}$. It is of combinatorial nature, and the initial sum over the intermediate spins $x$ has been fully performed. Notice that the combinatorial sum contains {\it a priori} $2^P$ terms, but only $2^{P-1}$ are actually different. This sum assigns different frequencies to the oscillations, since in particular it sums over the Regge actions of individual tetrahedra with all possible signs. That phenomenon has also been observed in \cite{dowdall-handlebodies}, where the authors looked at the asymptotics of the Ponzano-Regge model for 3d gravity on handlebodies and found a sum over `immersions' with different frequencies like here.

\section{Examples: Asymptotics of 15j-symbols with small and large angular momenta} \label{sec:15j}

A simple example is obviously the 9j-symbol with a small spin, treated in the first section. It corresponds to the case where there is a single sum over $\sigma=\pm1$, and $\omega^{(+)}_{k_1}\in [-2\pi,-\pi]$.

We now derive an asymptotic formula for the 15j-symbol with three small angular momenta, producing an alternative to the formula of \cite{yu15j}, and formulae when one, two, three and four small angular momenta. The latter are new to our knowledge. In particular, though the formulae derived in \cite{yu15j} look similar, they apply to different, non-equivalent choices of the small spins.

Using the notation of \cite{jucys}, we have
\begin{multline}
\wigner{j_1 & & j_2 & & j_3 & & j_4 & & j_5 \\
		& l_1 & & l_2 & & l_3 & & l_4 & & l_5 \\
		k_1 & & k_2 & & k_3 & & k_4 & & k_5}
=
\sum_x d_{x}(-1)^{R_5}
\wigner{j_1 & k_1 & x \\
		k_2 & j_2 & l_1}
\wigner{j_2 & k_2 & x \\
		k_3 & j_3 & l_2}
\wigner{j_{3} & k_{3} & x \\
		k_4 & j_4 & l_{3}}\\
\times \wigner{j_{4} & k_{4} & x \\
		k_5 & j_5 & l_{4}}
\wigner{j_5 & k_5 & x \\
		j_1 & k_1 & l_5},
\label{eq:15j}
\end{multline}
where $R_5 = \sum_{i=1}^5 j_i + k_i + l_i$. In what follows, we always assume that $j_1$ is small. Thus, every other small spin must be chosen among $l_2$, $l_3$ and $l_4$ (see subsection \ref{sbsec:hypotheses}).

\subsubsection{Four small angular momenta}

Assume that $j_1$, $l_2$, $l_3$ and $l_4$ are small. According to the notations of this section, we have
\begin{equation}
\setEdm = \{2,3,4\} \qquad \text{and} \qquad \setPR = \emptyset.
\end{equation}
It is easy to see from the derivation of \eqref{eq:general_formula} that if $\setPR = \emptyset$ then there is no sum over combinatorial sign assignments, $\{\sigma\} = \{0\}$. The corresponding tetrahedron \tet{\{\sigma\}} is flattened to a triangle with $\ph_{l_1,l_5} = \phi_1+\phi_5$. The reason is that in the decomposition \eqref{eq:15j}, all 6j-symbols have one small spin, so that no Ponzano-Regge formula gets involved. Consequently, there are no oscillations with some Regge action. Applying directly the general formula leads to
\begin{equation}
\{15j\} = \frac{ (-1)^{j_1+\mu} }{ \sqrt{ d_{l_1}d_{l_5}d_{j_2}d_{k_2}d_{j_3}d_{k_3}d_{j_4}d_{k_4} } }
\, \di_{\kappa_2\eta_2}^{(l_2)}(\ph_2)
\,\di_{\kappa_3\eta_3}^{(l_3)}(\ph_3)
\,\di_{\kappa_4\eta_4}^{(l_4)}(\ph_4)
\,\di_{\mu\nu}^{(j_1)}(\phi_1+\phi_5),
\label{eq:asympotic_15j_4_small_basic}
\end{equation}
with $\kappa_m \define k_{m+1}-k_m, \eta_m \define j_{m+1} - j_m, \mu \define j_2 - l_1$, and $\nu \define k_5 - l_5$.

\begin{figure}[!ht]
\centering
\includegraphics[scale=0.85]{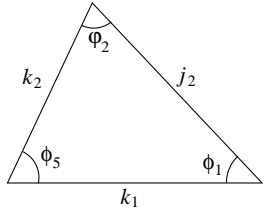}
\caption{Triangle defining the relevant geometric quantities for the asymptotics of the 15j-symbol with four small angular momenta.}
\label{fig:triangle_4_small_spins}
\end{figure}

That expression can be simplified. Since $l_2$, $l_3$ and $l_4$ are small, we have $j_2 \approx j_3 \approx j_4 \approx j_5$, and $k_2 \approx k_3 \approx k_4 \approx k_5$. Thus the triangles $(j_m,k_m,k_1)_{m\in\setEdm}$, $(k_1,j_2,k_2)$, and $(k_1,j_5,k_5)$ defined in Figure \ref{fig:angles_3_triangles} are identical. Consequently, we have $\ph_2 \approx \ph_3 \approx \ph_4 \approx \pi - \phi_1 - \phi_5$. The last (approximate) equality is simply due to the fact that the sum of the angles in a triangle is equal to $\pi$. The situation is illustrated in the Figure \ref{fig:triangle_4_small_spins}.

\noindent This finally leads to
\begin{equation}
\{15j\} \approx \frac{ 1 }{ d_{j_2}^2\, d_{k_2}^2 }
 \di_{\kappa_2\eta_2}^{(l_2)}(\ph_2)
\,\di_{\kappa_3\eta_3}^{(l_3)}(\ph_2)
\,\di_{\kappa_4\eta_4}^{(l_4)}(\ph_2)
\,\di_{\mu(-\nu)}^{(j_1)}(\ph_2).
\end{equation}
In \cite{yu15j}, Yu gives an expression for the asymptotics of the 15j-symbol with four small spins chosen among the five $l_i$. Although this configuration is different from ours, both formulae look quite similar.

\subsubsection{Three small angular momenta}

\begin{figure}[!ht]
\centering
\subfigure[\ \tet{4} which defines $V_4, S_{\rm R}^{(4)}$, with internal dihedral angle $\theta^{(4)}_{k_1}$.]{\includegraphics[scale=1]{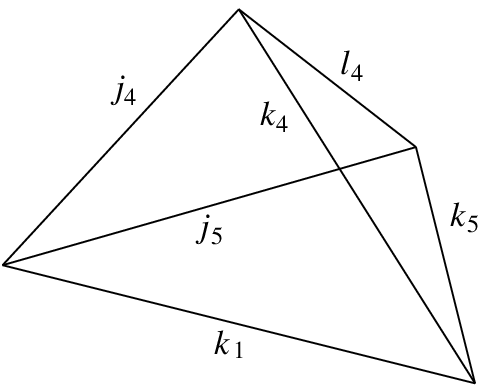}
\label{fig:tet_4_3_small_spins}}
\hspace{1.5cm}
\subfigure[\ \tet{(+)} which defines the geometric quantities $\ph_{j_4,k_5}^{(+)}$, $\theta_{j_4}^{(+)}$ and $\theta_{k_5}^{(+)}$.]{\includegraphics[scale=0.8]{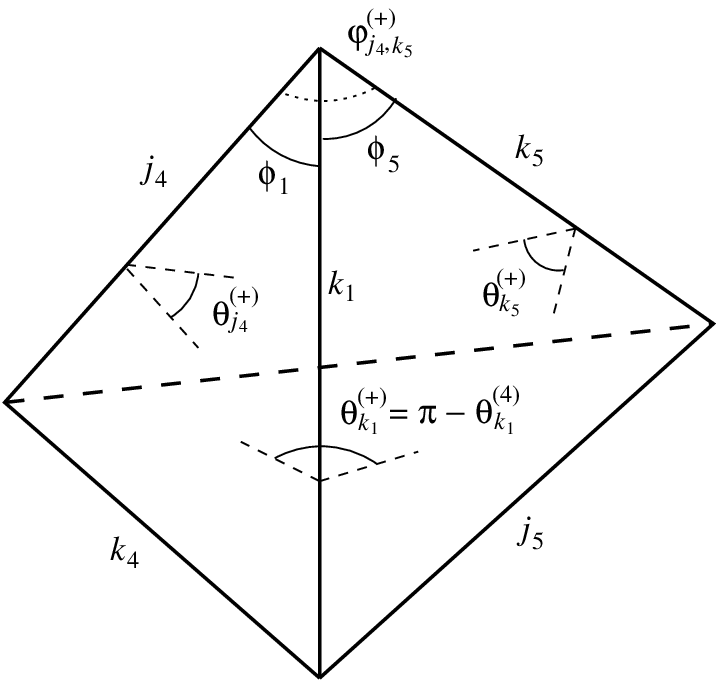}
\label{fig:tet_(+)_3_small_spins}}
\caption{Tetrahedra \tet{4} and \tet{(+)}. }

\end{figure}

We assume that $j_1$, $l_2$, and $l_3$ are small. The set of 6j-symbols on which we apply the Edmonds' and Ponzano-Regge formulae are
\begin{equation}
\setEdm = \{2,3\} \qquad \text{and} \qquad \setPR = \{4\}.
\end{equation}
There are $2$ sign configurations, $\sigma_4=\pm$, that we denote respectively $(+)$ and $(-)$. Their contribution in the combinatorial sum are equal, so we consider only the $(+)$ situation. The formula, once simplified, is
\begin{equation}
\{15j\}
\approx
\frac{ (-1)^{k_1+j_4+l_4+k_5+2j_1+\mu} }{ d_{j_2}\, d_{k_2}\ 12\pi V_4\,\sqrt{d_{j_2}\,d_{k_5}} }
\ \di_{\kappa_2 \eta_2}^{(l_2)}(\ph_2)
\,\di_{\kappa_3 \eta_3}^{(l_3)}(\ph_2)
\,\di_{\mu\nu}^{(j_1)} (\ph_{j_4,k_5}^{(+)})
\ \cos\pa{ S\e{R}^{(4)} + \frac{\pi}{4} - \mu \theta_{j_4}^{(+)} - \nu \theta_{k_5}^{(+)} + j_1 \pi},
\end{equation}
where $\ph_2 (\approx \ph_3 )$ is defined as usual; $V_4$ and $S\e{R}^{(4)}$ are the volume and the Regge action of \tet{4}, given in the Figure \ref{fig:tet_4_3_small_spins}; and the angles $\ph_{l_1,l_5}^{(+)}$, $\theta_{l_1}^{(+)}$ and $\theta_{l_5}^{(+)}$ belong to \tet{(+)}, represented in the Figure \ref{fig:tet_(+)_3_small_spins}. The latter is built by gluing the triangles $(k_1,l_1,k_2)$, $(k_1,l_5,j_5)$ with the dihedral angle $\theta^{(+)}_{k_1} \define \pi - \theta^{(4)}_{k_1}$, i.e. the external dihedral angle of \tet{4}.

The attentive reader may have noticed that the tetrahedron \tet{(+)} includes the triangle $(j_4,k_4,k_1)$ whereas for the general case (\tet{\{\sigma\}}, in Figure \ref{fig:tetrahedron_configuration_sigma}), we used $(l_1,k_2,k_1)$ instead. Both are actually equivalent since $j_4 \approx j_3 \approx j_2$ and $k_4 \approx k_3 \approx k_2$. Here, we have used that triangle to make contact with \tet{4}.

The tetrahedron \tet{(+)} is built out of the tetrahedron \tet{4} in exactly the same way the second tetrahedron (named \tet{2}) was built from the first tetrahedron in the case of the 9j-symbol with one small spin. Indeed, one flips one of the two triangles which share $k_1$, and set the external angle $\Theta^{(4)}_{k_1}$ as the new internal angle.

The underlying reason is that for the 9j-symbol with one small spin, just like for the 15j-symbol with three small spins and for any 3nj-symbol with $(n-2)$ small spins (chosen according to the hypotheses of Section \ref{sbsec:hypotheses}), one makes use of a single Ponzano-Regge asymptotics formula. Therefore only one tetrahedron with large angular momenta is involved with dihedral angle $\theta_{k_1}$ at $k_1$, and $\omega^{(+)}_{k_1} = 2\pi(n-2)+\theta_{k_1}$. Studying the range of $\omega^{(+)}_{k_1}$, one finds that the internal angle at $k_1$ in \tet{(+)} is always $\theta^{(+)}_{k_1} = \pi - \theta_{k_1}$, i.e. the external angle of the reference tetrahedron.
That means that the second tetrahedron is just built out of the first by the process we described.

In \cite{yu15j}, Yu has obtained an asymptotic formula for the 15j-symbol with $j_1, l_3, l_4$ being small. In that case, the tetrahedron of reference on which we have to apply the Ponzano-Regge formula is \tet{p=2} from the Figure \ref{fig:tet_regge}. However, Yu used a slightly different tetrahedron, which makes the formula a bit more complicated in the sense that one cannot express all the needed angles as angles from two tetrahedra only. Consequently the relevant Regge action is not the same as ours. Once the difference between the Regge actions is taken into account, it is possible to go from our formula to his' and to recover the geometric interpretation of his variables.

\subsubsection{Two small angular momenta}

We assume that $j_1$ and $l_2$ are small. The set of 6j-symbols on which we apply the Edmonds' and Ponzano-Regge formulae are
\begin{equation}
\setEdm = \{2\} \qquad \text{and} \qquad \setPR = \{3,4\}.
\end{equation}
There are $2^2=4$ sign configurations, $(\sigma_3,\sigma_4) = (++), (+-), (-+), (--)$. Thanks to the symmetry of the combinatorial sum, we can consider only the cases $(+\pm)$.

We now have two tetrahedra, \tet{3} and \tet{4}, with large angular momenta as in Figure \ref{fig:tet_regge} for $p=3,4$. To avoid distinguishing behaviors depending on their dihedral angles, we assume that they are nearly regular, so that their angles are close to $\theta_{\rm reg}= \arccos 1/3$. Also we assume without loss of generality that $\theta^{(3)}_{k_1} - \theta^{(4)}_{k_1}\geq0$.

\begin{figure}
\centering
\includegraphics[scale=0.85]{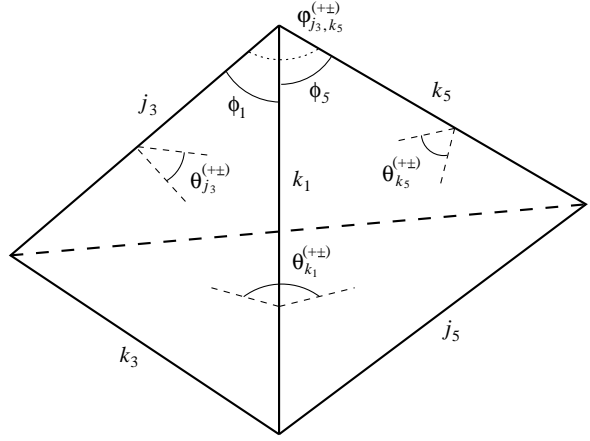}
\caption{Tetrahedra \tet{(++)}, \tet{(+-)} which are defined by the angles $\phi_1,\phi_5$ and $\theta^{(+\pm)}_{k_1} = \pi -(\theta^{(3)}_{k_1}\pm \theta^{(4)}_{k_1})$. They define in turn the angles $\ph^{(+\pm)}_{j_3,k_5}, \theta^{(+\pm)}_{j_3}$ and $\theta^{(+\pm)}_{k_5}$.}
\label{fig:tet_(++)_2_small_spins}
\end{figure}

The asymptotic formula is then
\begin{multline}
\{15j\} \approx
\frac{ (-1)^{j_3+l_3+j_4+k_4+l_4+k_5+j_1+\mu+2k_1} }{ 24\pi d_{j_3} \sqrt{ d_{k_3}\,d_{k_5}\,V_3\, V_4} }
\,\di_{\kappa_2 \eta_2}^{(l_2)}(\ph_2)
\left[
-\,\di_{\mu\nu}^{(j_1)} (\ph_{j_3,k_5}^{(++)})
\,\sin\pa{ S\e{R}^{(3)} + S\e{R}^{(4)} - \mu\theta_{j_3}^{(++)} - \nu\theta_{k_5}^{(++)} }\right. \\
+ \left.
(-1)^{2j_1}\,\di_{\mu\nu}^{(j_1)} (\ph_{j_3,k_5}^{(+-)})
\,\cos\pa{ S\e{R}^{(3)} - S\e{R}^{(4)} - \mu\theta_{j_3}^{(+-)} - \nu\theta_{k_5}^{(+-)} }
\right] .
\end{multline}
The volumes $V_3$, $V_4$ and Regge actions $S\e{R}^{(3)}$, $S\e{R}^{(3)}$ are associated with the tetrahedra \tet{3}, \tet{4}, see Figure \ref{fig:tet_regge}. The angles $\ph_{j_3,k_5}^{(+\pm)}$, $\theta_{j_3}^{(+\pm)}$ and $\theta_{k_5}^{(+\pm)}$ are defined in Figure \ref{fig:tet_(++)_2_small_spins} which pictures \tet{(+\pm)}.

Let us explain how to get the secondary tetrahedra \tet{(+\pm)}. Notice that \tet{3} and \tet{4} have a common triangle, $(k_1,j_4,k_4)$. Hence they can be glued together, in two different ways, either from the outside or one inside the other. We remove the common triangle $(k_1, j_4,k_4)$ so that the angle at $k_1$ between $(k_1,j_3,k_3)$ and $(k_1,j_5,k_5)$ is either $\theta^{(3)}_{k_1}+\theta^{(4)}_{k_1}$, or $\theta^{(3)}_{k_1}-\theta^{(4)}_{k_1}$.

The generic study tells us to consider the angle $\omega^{(+\pm)}_{k_1} \define 6\pi -\Theta^{(3)}_{k_1} \mp \Theta^{(4)}_{k_1}$. One gets
\be
\omega^{(+\pm)}_{k_1} = \theta^{(3)}_{k_1} \pm \theta^{(4)}_{k_1} -(1\mp 1)\pi \pmod{4\pi},
\ee
which implies $\omega^{(++)}_{k_1}\in[0,\pi]$ and $\omega^{(+-)}_{k_1}\in[-2\pi,-\pi]$. As a conclusion, the tetrahedra \tet{(+\pm)} are built by flipping the triangle $(k_1,j_5,k_5)$ so that $k_5$ meet $j_3$ while $j_5$ meet $k_3$. Then from those five lengths, get a tetrahedron by setting as the new dihedral angle at $k_1$
\be
\theta^{(+\pm)}_{k_1} = \pi - \bigl(\theta^{(3)}_{k_1}\pm \theta^{(4)}_{k_1}\bigr),
\ee
like in Figure \ref{fig:tet_(++)_2_small_spins}. 

In \cite{yu15j} a formula for the 15j-symbol with two small angular momenta is given, which are $l_3, l_4$. That configuration is not the same as here, but the formulae look quite similar. A notable difference is the denominator of the amplitude which is here the product of the volumes of the two relevant tetrahedra, while it is more complicated in \cite{yu15j}.

\subsubsection{One small angular momentum}

We assume that only $j_1$ is small. The set of 6j-symbols on which we apply the Edmonds' and Ponzano-Regge formulae are
\begin{equation}
\setEdm = \emptyset \qquad \text{and} \qquad \setPR = \{2,3,4\}.
\end{equation}
There are $2^3=8$ sign configurations $(\sigma_2,\sigma_3,\sigma_4)$. Thanks to the symmetry of the combinatorial sum, we can consider only $(+++)$ and $(++-),(+-+),(-++)$.

\begin{figure}[!h]
\centering
\includegraphics[scale=0.9]{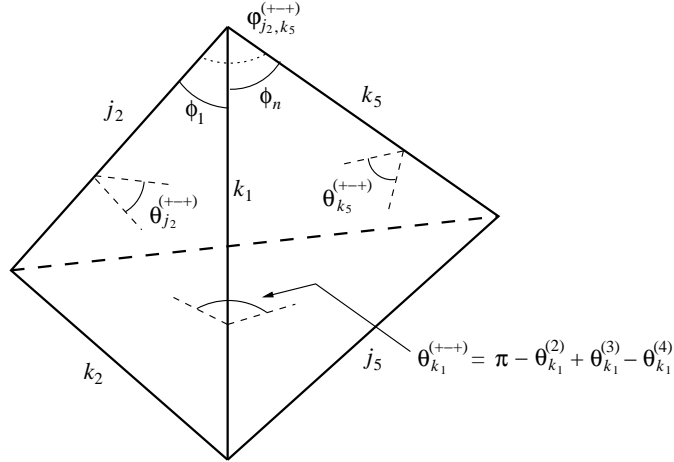}
\caption{Tetrahedron \tet{(+-+)} which defines the geometric quantities $\ph_{j_2,k_5}^{(+-+)}$, $\theta_{j_2}^{(+-+)}$ and $\theta_{k_5}^{(+-+)}$, given $\phi_1,\phi_5$ and $\theta^{(+-+)}_{k_1}$.}
\label{fig:tet_(+-+)_1_small_spin}
\end{figure}

For simplicity, we assume that the three tetrahedra associated to $\setPR$ are almost regular. We get
\begin{multline}
\{15j\}
\approx
\frac{ (-1)^{(j_2+l_2+j_3) + (k_3+l_3+k_4) + j_4+l_4+k_5-k_1+\mu} }{ 48\pi \sqrt{12\pi \,d_{j_2} d_{k_2} V_2 V_3 V_4} }\\
\Bigl[\,
\di_{\mu\nu}^{(j_1)} (\ph_{j_2,k_5}^{(++-)})\,
\cos\pa{ S\e{R}^{(2)} + S\e{R}^{(3)} - S\e{R}^{(4)} + \frac{\pi}{4} - \mu\theta_{j_2}^{(++-)} - \nu\theta_{k_5}^{(++-)} + \pi j_1}
\\ + 
\di_{\mu\nu}^{(j_1)} (\ph_{j_2,k_5}^{(+-+)})\,
\cos\pa{ S\e{R}^{(2)} - S\e{R}^{(3)} + S\e{R}^{(4)} + \frac{\pi}{4} - \mu\theta_{j_2}^{(+-+)} - \nu\theta_{k_5}^{(+-+)} + \pi j_1}
\\ + 
\di_{\mu\nu}^{(j_1)} (\ph_{j_2,k_5}^{(-++)})\,
\cos\pa{ - S\e{R}^{(2)} + S\e{R}^{(3)} + S\e{R}^{(4)} + \frac{\pi}{4} - \mu\theta_{j_2}^{(-++)} - \nu\theta_{k_5}^{(-++)} + \pi j_1}
\\ + 
\di_{\mu\nu}^{(j_1)} (\ph_{j_2,k_5}^{(+++)})\,
\cos\,\Bigl( S\e{R}^{(2)} + S\e{R}^{(3)} + S\e{R}^{(4)} + \frac{3\pi}{4} + \mu\theta_{j_2}^{(+++)} + \nu\theta_{k_5}^{(+++)} + \pi j_1 \Bigr)
\Bigr].
\end{multline}
The volumes $V_3, V_4, V_5$ and Regge actions $S\e{R}^{(2)}$, $S\e{R}^{(3)}$, $S\e{R}^{(4)}$ are associated with the tetrahedra \tet{2}, \tet{3}, \tet{4}, respectively, see the Figure \ref{fig:tet_regge} for $p=2,3,4$. The angles $\ph_{j_2,k_5}^{(\pm\pm\pm)}$, $\theta_{j_2}^{(\pm\pm\pm)}$ and $\theta_{k_5}^{(\pm\pm\pm)}$ belong to the tetrahedra \tet{(\pm\pm\pm)}. They are defined by the five spins $k_1, j_2,k_2,j_5,k_5$ which form two triangles glued along $k_1$ like in Figure \ref{fig:tet_(+-+)_1_small_spin}. To complete their definition, we set the dihedral angle at $k_1$ to be
\begin{gather}
\nonumber \theta^{(+++)}_{k_1} = \theta^{(2)}_{k_1} +\theta^{(3)}_{k_1} +\theta^{(4)}_{k_1} -\pi,\\
\nonumber \theta^{(++-)}_{k_1} = \pi -\theta^{(2)}_{k_1} - \theta^{(3)}_{k_1} + \theta^{(4)}_{k_1},\\
\nonumber \theta^{(+-+)}_{k_1} = \pi -\theta^{(2)}_{k_1} + \theta^{(3)}_{k_1} - \theta^{(4)}_{k_1},\\
\theta^{(-++)}_{k_1} = \pi +\theta^{(2)}_{k_1} - \theta^{(3)}_{k_1} - \theta^{(4)}_{k_1}.
\label{eq:angle1small}
\end{gather}
Note that since \tet{2}, \tet{3}, \tet{4} are close to being regular the above angle indeed lie in $[0,\pi]$.

The intuitive way of building those tetrahedra is by first gluing \tet{2}, \tet{3} along their common triangle $(k_1,j_3,k_3)$. There are two different ways, either on the outside, or one inside the other. Then, one adds \tet{4} by gluing it along the triangle $(k_1,j_4,k_4)$, and there are here again two different ways to do so. Finally we focus on the triangles $(k_1,j_2,k_2)$ and $(k_1,j_5,k_5)$ which can be completed to a tetrahedron, and we consider the tetrahedron which complements that one when drawing the parallelogram spanned by $(j_2,k_2)$. The final internal angle is given by \eqref{eq:angle1small}, depending on the chosen way to glue the three initial tetrahedra. As an example, we give \tet{(+-+)} in Figure \ref{fig:tet_(+-+)_1_small_spin}.

\section{Conclusion}

We have shown that the result of \cite{Yu_and_Littlejohn} can be recovered by direct application of the Ponzano-Regge formula (together with Edmonds' formula), making clear the way the asymptotic information is encoded into the geometry of a tetrahedron. Moreover, the conditions of applicability of our method have been given and we derived new, explicit formulae for asymptotics of arbitrary Wigner symbols with some large and small angular momenta. Our method is simpler than that of \cite{Yu_and_Littlejohn}, as far as Wigner symbols are concerned, and provides us with complementary results.

Some directions for future research are proposed in the Conclusion of \cite{Yu_and_Littlejohn} and are definitely of interest.

We would like to mention in addition that beyond the asymptotic formulae for specific symbols, the regime where some spins remain small should deserve attention, because the geometric content is simply contained in tetrahedra. In particular, it may be investigated in terms of recurrence relations. As argued in \cite{recurrence-paper}, the latter provides a preferred way to encode the geometric properties of classical spin networks. It was further shown in \cite{3d-wdw}, in the case of the 6j-symbol, that it is possible to derive those recurrences as quantum constraints (Wheeler-DeWitt equations) coming from three-dimensional gravity. The same result holds in the four-dimensional context \cite{semiclass-paper} using a Hamiltonian from topological field theory. Hence, it would be interesting to look at the asymptotics of those relations with small and large spins, and to look for a similar regime where the asymptotic geometry is described in terms of 4-simplexes instead of tetrahedra.

\section*{Acknowledgements}

P.F. acknowledges the Summer Program of Perimeter Institute for undergraduate students, without which this project could have been completed.

Research at Perimeter Institute is supported by the Government of Canada through Industry Canada and by the Province of Ontario through the Ministry of Research and Innovation.


\end{document}